\documentclass[12pt]{article}

\input{tcilatex}

\begin{document}

\begin{center}
\textbf{Formation of Languages; Equality, Hierarchy and Teachers}

\c{C}a\u{g}lar Tuncay

Department of Physics, Middle East Technical University

06531 Ankara, Turkey

caglart@metu.edu.tr
\end{center}

\textbf{Abstract: }A quantitative method is suggested, where meanings of
words, and grammatic rules about these, of a vocabulary are represented by
real numbers. People meet randomly, and average their vocabularies if they
are equal; otherwise they either copy from higher hierarchy or stay idle.
Presence of teachers broadcasting the same (but arbitrarily chosen)
vocabulary leads the language formations to converge more quickly.

\textbf{Introduction:} Within the emerging physics literature on languages [%
\textbf{1-12}], birth of a language may be observed as a scarcely studied
issue. In our opinion, the subject is important for researches on
language competition, since quickly developed languages may have more chance
to survive and to spread. In the present contribution, effect of inequality
on the speed of originating a language is studied, where some social
agents (hierarchical people, teachers) play the role of nucleation centers for
clustering of words, meanings, and grammatic rules, etc. We present a
quantitative model, where each subentry of a vocabulary is represented by a
real number, and so are the words. Model is given in the following section;
applications and results are displayed in next one. Last section is devoted
for discussion and conclusion.

\bigskip
\textbf{Model:} We have a society composed of $N$ adults. Each
person $k$ has a vocabulary of $M$ words ($w_{ki}$, $k\leq N$, $i\leq M$). For a
word there exist many related items as meanings, rules for plural forms,
adverb forms, tenses, prefixes, suffixes, etc. In real life, many words have
a diversity of such peculiarities, which are not all easy to learn and to
remember; since their meanings may be close to each other, as
\textquotedblleft dictionary\textquotedblright\ and \textquotedblleft
vocabulary\textquotedblright . Pronunciations may be similar too; as
\textquotedblleft a head\textquotedblright , and \textquotedblleft
ahead\textquotedblright . Also; as, \textquotedblleft
night\textquotedblright , \textquotedblleft knight\textquotedblright , and
\textquotedblleft knife\textquotedblright . Such variations are symbolized
by five different subentries ($j$) at most. So we take $1\leq j\leq j_{max}$ 
for every word $w$ and for each $j$ we assign a representative real number $r$.
Therefore our words are sets of up to five real numbers:

$$\noindent w_{ki}={\{}r_{kij}{\}} . \eqno(1)$$

The maximum number $j_{max}$ of subentries $r$ is also determined randomly
between 1 and 5, independently for any word $w$. Clearly, $r_{kij}=0$ 
($w_{ki}= {\{}0,0,0,0,0{\}}$) corresponds to an unknown meaning (word) in the
vocabulary of the adult $k$.

Initially there is no consensus about a common vocabulary, but the
consensus may be set through several processes described in the following
subsections, and the values for initial $r_{kij}$'s in Eq. (2) must be
changed into time dependent ones, i.e. $r_{kij}(t)$.

Evolution of the language spoken by any adult may be described by

$$L_{k}(t)=\sum_{i=1}^{M}\; \sum_{j=1}^{j_{max}} \; r_{kij}(t), \eqno(2)$$
wherex, $L_{k}$ varies from person to person, especially at the
beginning of the formation period, and this fluctuation fades down with time
since $L_{k}\rightarrow L$, if convergence occurs.

Eq. (2) may be summed over the members ($k$) of the society to consider all
the vocabularies present at time $t$:

$$D(t)=\sum_{k}^{N}L_{k}(t) . \eqno(3)$$
As we observed, $D(t)-D(t-1)$ is a significant quantity within the present
formalism, and we represent it by $V(t)$:

$$V(t)=D(t)-D(t-1) . \eqno(4)$$

As $t\rightarrow \infty, \; D(t)$ is expected to converge to its limit 
$D(t\rightarrow \infty )$, and $L_{k}(t)$ to some $L$, and $V(t)$ to zero.
Then, the language $L=L_k(t\rightarrow \infty )$ may be evaluated as
established. Minor fluctuations within $D(t)$ about $D(t\rightarrow \infty )$, 
and these within $V(t)$ about zero may be attributed to misuses due to
lack of individual memories to remember all the relevant meanings, and
rules, etc.

\underline{\textit{Initiation}}$:$ We assign random real numbers for initial
values of $r_{kij}$, with $0\leq r_{kij}<1$, where $k\leq N,i\leq M$, and $%
j\leq j_{max}(k,i)$.

\underline{\textit{Evolution}}$:$ Once the initial vocabularies are set, we
assume that two members ($k$, and $k\mbox{'})$ meet randomly at a time $t$.

In the simplest case of no inequality in status (\underline{\textit{equality}}),
they average [\textbf{13-14}] subentries ($r_{kij})$ in their
vocabularies, and share the new ones:

$$r_{kij}(t)=(r_{kij}(t-1)+r_{k\mbox{'}ij}(t-1))/2=r_{k\mbox{'}ij}(t),\eqno(5)$$
and the language spoken by each adult ($k$) becomes:

$$L_{k}(t)=L_{k^{\prime }}(t)=(L_{k}(t-1)+L_{k^{\prime }}(t-1))/2 . \eqno(6)$$

As interaction tours (time $t$) advance, r$_{kij}(t\rightarrow \infty )=1/2$
independent of the subindices. We have $D(t)=D(0)$ and $V(t)=V(0)=0$, for
all $t$, since $L_{k}(t)+L_{k^{\prime }}(t)=L_{k}(t-1)+L_{k^{\prime }}(t-1)$%
, due to Eq. (6).

We incorporate \underline {\textit{inequality}} into the society, by
assigning some rank to adults in terms of real numbers (greater than or
equal to zero, and less than one) determined randomly. Yet, any two adults
will be considered as equivalent if their ranks are close to each other by a
given $\Delta $, and each member will average her vocabulary with the other,
Eq. (5) and (6). Otherwise, the one with lower rank (obeying) will copy
down the vocabulary of the other (commander) and take it as her new
vocabulary, till another possible meet with any adult occurs. In this case
convergence (formation) of the language may be speeded up under certain
conditions, as studied within the following section.

Furthermore, we may assume more stringent inequality: Some hierarchy (all,
with rank of unity) broadcast the same (yet arbitrarily selected) vocabulary
to the society, from the beginning on. We call them teachers. They will not
change their common vocabulary and due to their ultimate rank, they will not
average their vocabularies with anyone. Some other hierarchical people (within
a given limit of $\Delta )$ may average their vocabularies after they
discuss with teachers. And the rest copies down from all those who have
higher ranks by $\Delta $.

\bigskip
\textbf{Applications and Results:} In this section we will first consider
uniqueness within society. Later, by assigning to each individual a random
real number (rank; greater than or equal to zero, and less than one) we will
establish hierarchy. And finally, we will incorporate some teachers with
ultimate rank of unity into society.

We handled \underline{\textit{equality}} within adults by assuming an
averaging process for the words ($w_{ki}$ of Eqs. (1), (2), and (5)), and
the meanings ($r_{kij}$ of Eq. (1)) [\textbf{13-14}]. Evolution of $%
r_{kij}(t)$, for a randomly selected $j$ is displayed in Figure 1, where
adults ($N=500$) are all equal and only arbitrarily chosen hundred
adults are displayed. Each adult had her own initial randomly selected
meanings ($r_{kij}(0)$) as used by herself and suggested to the society.
Whenever any two of the adults randomly meet, they obey Eq. (5); each
interacts equally with the other and averages her vocabulary. $D(t)=D(0)$
and $V(t)=V(0)=0$, for all $t$, since $D(t)$ of Eq. (3) does not change
during interactions $L_{k}(t)+L_{k^{\prime }}(t)=L_{k}(t-1)+L_{k^{\prime
}}(t-1)$, Eq. (6)). Corresponding probability density function (PDF) for $%
r_{kij}(t)$ (with $N=500$, $M=100$ and $j\leq j_{max}$) is a
delta function, i.e., PDF$(V)=\delta (0)$ (inset, Fig. 1.).

We incorporate \underline {\textit{inequality}} into the society, by
assigning some rank to adults in terms of real numbers (greater than or
equal to zero, and less than one) determined randomly. Yet, any two adults
will be considered as equivalent if their ranks are close to each other by a
given $\Delta $. Under the present condition, each member will average her
vocabulary with the other, Eq. (3). Otherwise, the one with lower rank will
copy down the vocabulary of the other and take it as her new vocabulary,
till another possible meet with any adult occurs.

For small $\Delta $, almost everybody (except the top of hierarchy with rank 
$1-\Delta $) may copy from others, and almost everybody (except the bottom of
hierarchy with rank $\Delta )$ may be copied by others. Within this content, the
averaging process between equals is ignored within the society ($N$). On the
other hand if $0.5<\Delta $, only the top of hierarchy with rank $1-\Delta $
will be copied by the bottom of that with rank $\Delta $, and more than half
of the society will average. Clearly, averaging process will dominate as $0.5
\ll \Delta \rightarrow 1.0$; therefore this regime implies more freedom
and more discussion. $\Delta =1.0$ case corresponds to equality of all the
adults.

Evolutions of $r_{kij}(t)$ with various $N$, and $M$ as designated in the
figure captions and $j\leq j_{max}$, and $D(t)$, and $V(t)$ are displayed, in
Figures 2a, and b, and c, respectively, where $\Delta =0.2$ for all. One
may remark that, discussing and averaging mechanism between (close) equals
(by $\Delta $), or copying from the vocabulary of some higher rank people
causes the language to converge, yet convergence is very small for $\Delta
\sim 0$, and speeds up as $\Delta \rightarrow 1$. For $\Delta \sim 0$, all
the society speaks ultimately the language of the one with highest rank
which is very close to unity.

\underline{\textit{Teachers}}: Figure 3a displays evolution of $r_{kij}(t)$, 
with randomly selected number of meanings ($j\leq j_{max}$), where
the meanings of words belong to language of arbitrarily chosen hundred
adults out of $N=5000$ adults. Please note the horizontal
limiting line representing the language broadcasted by teachers. The greater the
distance from this line is, the greater is the needed effort to learn the
language. Figure 3b displays $D(t)$, and Figure 3c displays $V(t)$,
with $\Delta =0.2$ and $\tau =0.2$ in all, where $\tau $ designates the
number of teachers per population of the society ($N$). Please note the
rapid convergence in $D(t)$ and $V(t)$.

In Figure 3c there exist three behaviors in $V(t)$: For $t\sim 0$ region
we have comparably big fluctuations; for $t\rightarrow \infty $ we have very
small fluctuations, both about zero. And in between we have exponential
decay. Initial fluctuations originate from randomness, and the number of
equilibrated ones may be increased by increasing the number of tours (and
also, precision of real numbers in the utilized software). So, the
characteristic regime is the intermediate one and exponential decay implies
that the envelope function for $D(t)$ (which passes through local maxima and
minima) is also an exponentially decaying one. (We had observed similar
exponential decays within our computations on opinion dynamics. [\textbf{16}%
]) The pronounced threefold behavior is reflected in PDF's in Figure 3c;
where, the horizontal axis is for $V^{2}$, and the perpendicular one is
logarithmic.

Small-speed regimes in PDF's of Figure 3d correspond to $t\rightarrow
\infty $ region in $V(t)$, which may be ignored totally. Please note that 
PDF$(V)$ (and PDF$(V^{2})$) goes to $\delta $, as $\Delta $ approaches unity
and high speed wing tips in PDF's are coming from $t\sim 0$ region in $V(t)$%
, where randomness is dominant. Teachers shape the intermediate region, and
due to them we have the exponential convergence in $D(t)$. And one new
language emerges, which is spoken by the majority of adults, and will be
learned by children.

\bigskip
\textbf{Discussion and Conclusion:} Clearly, increasing the number of
teachers (and $\tau $) increases rates of exponential decays in $V(t)$ and
$D(t)$: There will be more chance to check personal vocabularies, and
number of ordinary adults will be lowered. Big differences between the real
numbers associated to entries of the broadcasted common vocabulary and those
to initial settings may be considered as a kind of measure for difficulty to
learn the relevant language, since more interaction tours will be needed
for averaging before the personal vocabularies approach the broadcasted
one. If the equilibrium level of $D(t)$ is far from the initial one, then the
emerging language may be considered as a tough one to learn. (We run the
case, with $0.9\leq r_{kij}<1.0$, and $0.0\leq r_{kij}<0.1$ (Eq. 3) for the
teachers' vocabulary many times and verified the last remark in all.)

We run also the case, where each teacher broadcasted (keeping her ultimate
rank) a different vocabulary, rather than a common one. This case
corresponds to a richer language. And we obtained still, but rather slower,
exponential decays. In our opinion, this result agrees well with the
reality that those languages involving more words and grammatic rules are
harder to learn than those with less words and rules. In any case,
presence of nuclei speeds up clustering of words and rules; and the
relevant language emerges quickly. So, when a group of people immigrate to a
new society, and if they gain rank (power) they may broadcast their language
to the present society, which may be considered as one of the possible
mechanism to spread languages besides colonization, conquest, etc. As a
final remark it may be stated that we varied the number of words (upper
limit within the sum of Eq. (2)) and the number of adults ($N$) within the
society from 10, 100 to 1000, 5000, all respectively and obtained similar
results. As the numbers decreased, fluctuations increased; yet, the envelope
of $D(t)$ always came out as exponential.

\bigskip
\textbf{Acknowledgement}

The author is thankful to Dietrich Stauffer for his inspirations and
friendly discussions and corrections, and informing about the references [%
\textbf{12-15}].

\bigskip
\textbf{Reference List}
\parindent 0pt

[1] D.M. Abrams and S.H. Strogatz, Nature \textbf{424}, 900 (2003).

[2] M. Patriarca and T. Leppanen, Physica \textbf{A 338}, 296 (2004).

[3] J.Mira and A. Paredes, Europhys. Lett. \textbf{69}, 1031 (2005).

[4] C. Schulze and D. Stauffer, Int'l. J. Mod. Phys. C \textbf{16}, 781
(2005).

[5] K. Kosmidis, J.M. Halley and P. Argyrakis, Physica A \textbf{353}, 595
(2005).

[6] J.P. Pinasco and L. Romanelli, Physica A 361, 355 (2006).

[7] V. Schwammle, Int'l. J. Mod. Phys. C \textbf{16}, 1519 (2005).

[8] V.M. de Oliveira, M.A.F. Gomes and I.R. Tsang, Physica A \textbf{361},
361 (2006).

[9] V.M. de Oliveira, P.R.A. Campos, M.A.F. Gomes and I.R. Tsang, Physica A 
\textbf{368}, 257 (2006).

[10] A. Baronchelli, M. Felici, E. Caglioti, V. Loreto, L. Steels, J.
Stat.Mech., P06014 (2006).

[11] D. Stauffer, X. Castello, V.M. Eguiluz and M. San Miguel, e-print
physics/0603042 at \underline{www.arXiv.org}. Will be published in Physica A.

[12] P.M.C. de Oliveira, D. Stauffer, F.W.S. Lima, A.O. Sousa, C. Schulze,
and S. Moss de Oliveira, ``Bit-strings and other modifications of Viviane model 
for language competition'', preprint for Physica A.

[13] G. Deffuant et al., J. Artificial Societies and Social Simulation, vol.
5, no. 4 (2002).  URL: http://jasss.soc.surrey.ac.uk.

[14] P. Assmann, Int'l. J. Mod. Phys. C \textbf{15}, 1439 (2004).

[15] D. Stauffer, Computing in Science and Engineering \textbf{5}, 71 (2003).

[16] \c{C}. Tuncay, ``Opinion Dynamics for number of transactions and price,
a trader based model'' e-print/physics/0604179, at \underline{www.arXiv.org}. 
Will be published in Int'l J. Modern Physics C, (2006).
\bigskip

\textbf{FIGURES}

\textbf{Figure 1 }Evolution of $\sum_{i}^{M}r_{kij}(t)$, for three adults,
where $j$ is arbitrary with $j\leq j_{max}$, and $M=100$, for 
$N=500$. Inset shows PDF for time rate of change of $r_{kij}(t)$.

\textbf{Figure 2a} Evolution of $\sum_{i}^{M}r_{kij}(t)$, for three
adults, where $j$ is arbitrary and $M =100$, 
for $N=1000$, where $\Delta =0.2$.

\textbf{Figure 2b} Evolution of $D(t)$ with $M=300$%
\textbf{, }$N=5000$, for $\Delta =0.2$.

\textbf{Figure 2c} Evolution of V(t) with $M=300$%
\textbf{, }$N=5000$, for $\Delta =0.2$.

\textbf{Figure 3a} Evolution of $\sum_{i}^{M}r_{kij}(t)$, for
three adults, where $j$ is arbitrary and $M =300$, 
for $N=1000$, with $\Delta =0.2$, $\tau =0.2$.

\textbf{Figure 3b} $D(t)$ with $\Delta =0.2$ and $\tau =0.2$.
Please notify the rapid convergence.

\textbf{Figure 3c} $V(t)$ with $\Delta =0.2$ and $\tau =0.2$.
Please notify the rapid convergence in $V(t)$. Perpendicular axis for $V(t)$
is logarithmic. The inset shows PDF for the given $V(t)$.

\textbf{Figure 3d} PDF$(V)$ for $\Delta =0.2$, and various $\tau 
$, where the horizontal axis is $V^{2}$, and the perpendicular one is
logarithmic.

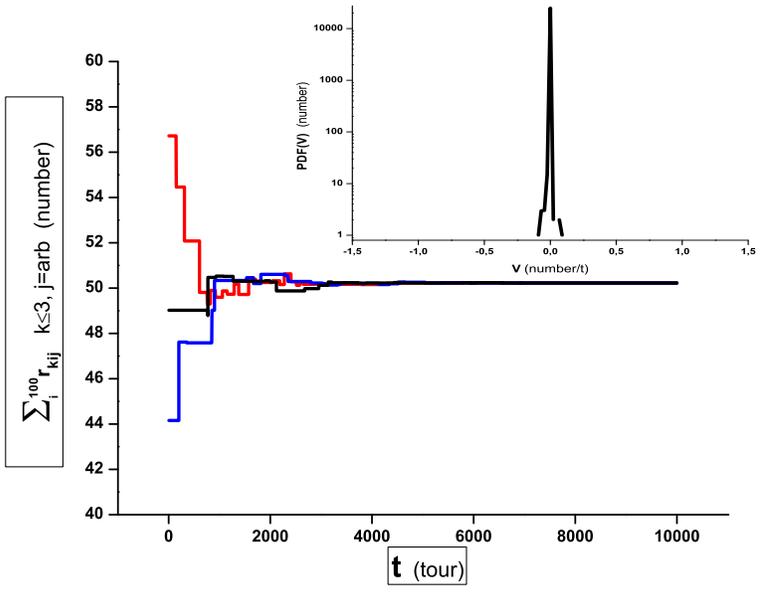
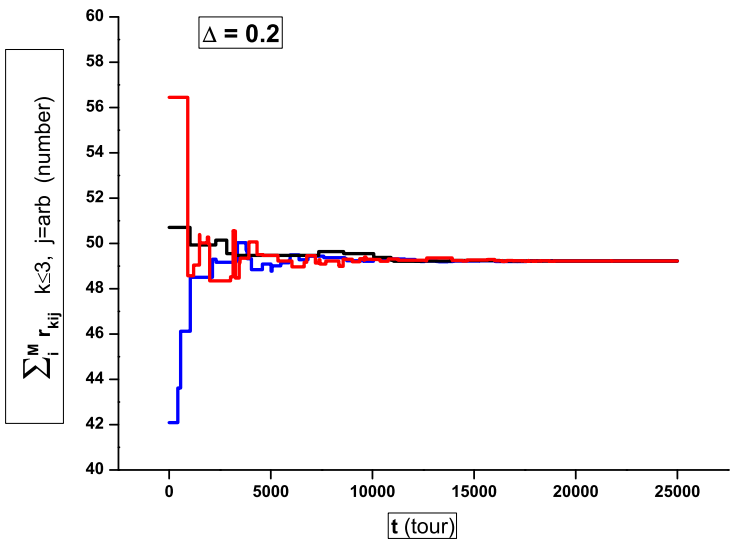
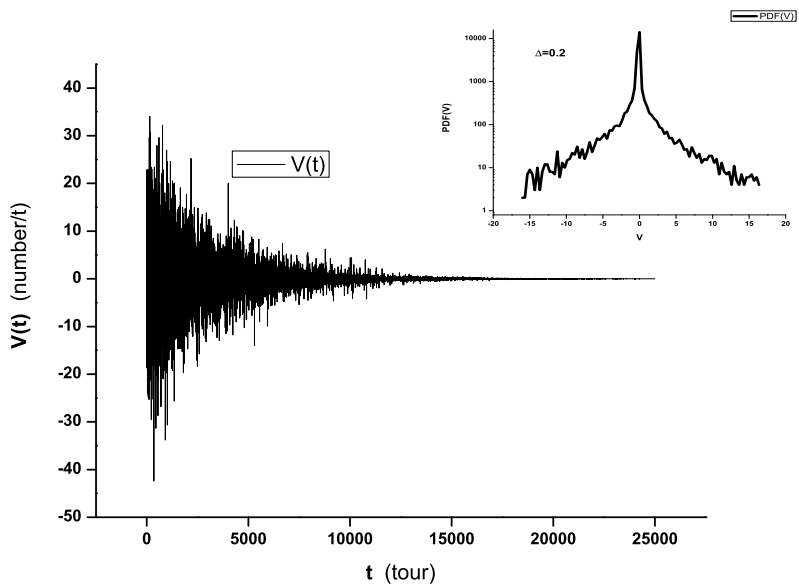
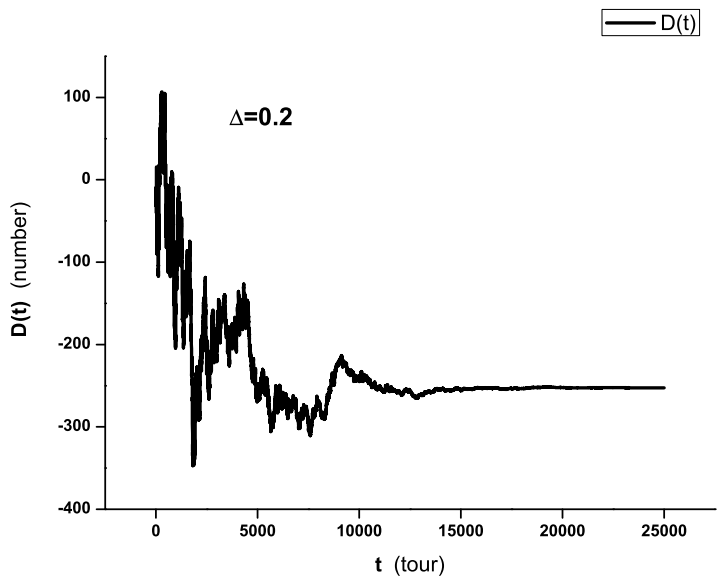
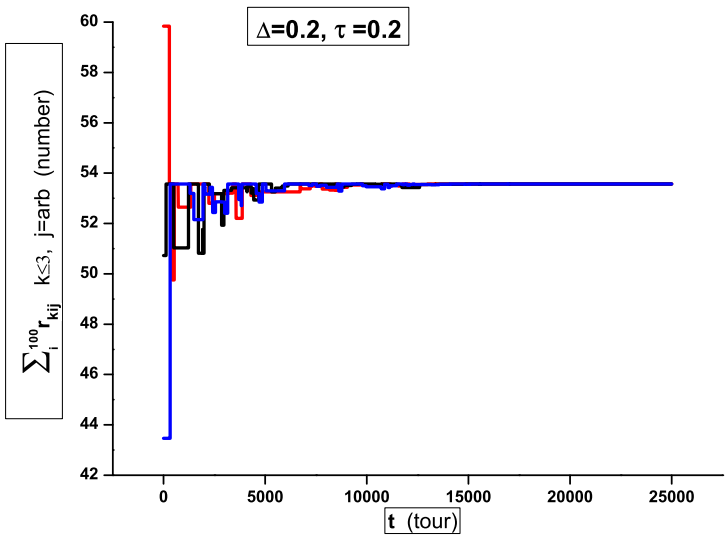
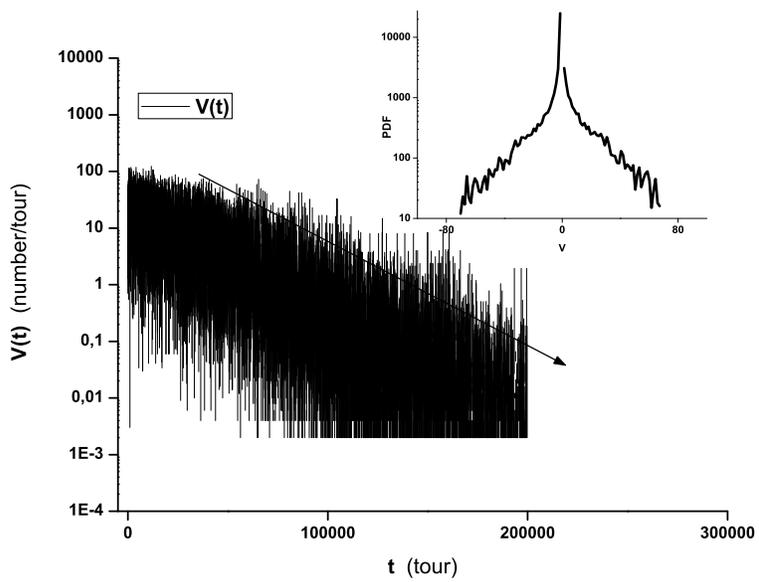
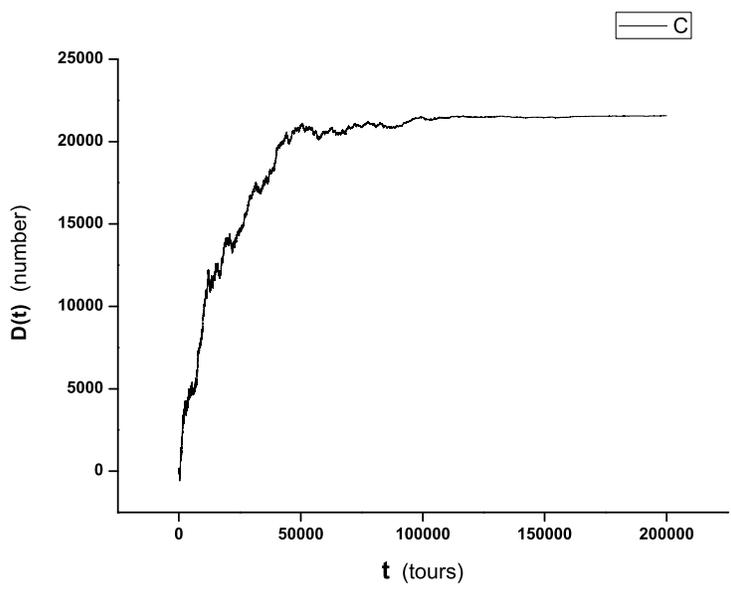
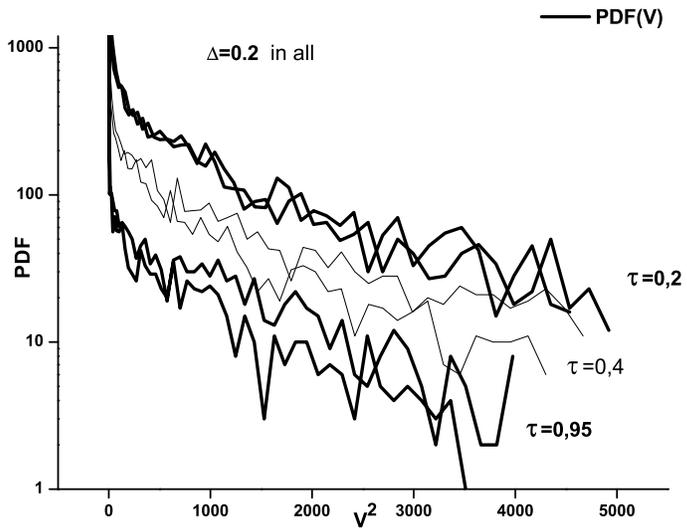
\end{document}